\documentstyle[aps,prl,epsfig,preprint]{revtex}
\begin{document}

{\bf Comment on   ``Negative Heat Capacity for a Cluster of 
147 Sodium Atoms" by Schmidt et al.}

\bigskip

In a recent letter \cite{schmidt}, Schmidt et al. analyze a 
region where the specific heat of a $Na_{147}^+$ cluster becomes negative. 
In Boltzmann-Gibbs (BG) statistical mechanics the specific heat  
can never be negative in the canonical ensemble.
However,
as the authors are well aware,
no such property is valid in general for the microcanonical ensemble. 
This is indeed the case of a variety of physical situations ranging
from self-gravitating systems to melting atomic clusters and
fragmenting nuclei (see \cite{schmidt} and references therein).
An important feature that all such systems have in common 
is thermodynamic {\it nonextensivity}, reflected,  
as also mentioned in \cite{schmidt}, in the fact that 
the energy associated with $N$ particles is not proportional to $N$. 
Nonextensivity occurs everytime {\it the range
of the interactions is not negligible compared 
to the (linear) size of the system}. Two important 
physical examples are:

\noindent
-- {\it short-range interactions in small systems }

\noindent
-- {\it long-range interactions} in systems of any size
\cite{fisher,tsallis,stefano}.
\\
Now, if the system is isolated, nonextensive, 
in an equilibrium-like state, and exhibits, for some energy range, 
a negative specific heat, then not one but at least two 
possibilities must be considered:

1) the system is in its thermodynamically
stable state and then the BG microcanonical ensemble is to be used, 
as preconised in \cite{gross}  
and done in \cite{schmidt}.

2) the system is in some kind 
of dynamic {\it metastable state}, and then some other
thermostatistical approach might well be necessary.
One such alternative description is nonextensive statistical 
mechanics \cite{tsallis}, which has proved to be useful in a 
variety of complex situations (e.g., fully developed 
turbulence \cite{turbulence}, electron-positron 
annihilation producing hadronic 
jets \cite{bediaga}, motion of {\it Hydra viridissima} \cite{arpita}).

The metaequilibrium state to which we make reference as a second 
possibility has indeed been observed in magnetic-like Hamiltonians 
such as arrays
of long-range coupled planar rotators (the so called HMF \cite{HMF}
and its generalization, the $\alpha-XY$ model \cite{alfaxy}), among 
others, including fluid-like systems \cite{borges}.
To illustrate this point, we report in fig. 1 a numerical simulation 
typical of the relaxation process in the HMF model. 
As a function of time, two plateaux are observed 
for the average kinetic energy per particle
and the BG formalism is violated for the first plateau 
(panel a), where we find a region of negative 
specific heat (panel c)\cite{nonmaxwell}.  
More precisely, the velocity distribution is
well described by the standard Maxwellian in the late plateau (panel b), 
whereas for the early plateau it appears to be described by the 
type of distribution emerging within nonextensive statistical 
mechanics \cite{nonmaxwell}. 
Similar results and anomalies have 
been recently described in other models, including first order 
phase transitions \cite{sota,posch,hmf2}.

The relevance of long range interactions in sodium clusters 
and similar atomic liquids is well known \cite{wales}. 
Consequently metastable states could very well occur in relatively 
small clusters
such as those studied in \cite{schmidt}, 
as in various colloidal glasses \cite{colloid} and 
other clusterized systems \cite{bene}.
For example, in colloidal glasses it is common belief that the
structural arrest at the glass transition is due to the increasingly
size of strongly cooperating sets of atoms \cite{colloid}.
Under these circumstances, and without further
experimental evidence, application of BG equilibrium 
statistical mechanics in \cite{schmidt} might have no support.  
Since the analysis of the authors only applies to one among 
at least two physically important possibilities, 
the theoretical interpretation  they draw from their  
experiments \cite{schmidt} 
is not necessarily true.

Useful remarks from J. Pacheco, K.A. Dawson and A. Lawlor 
are gratefully acknowledged. Partial support from 
FCT (Portuguese agency), CNPq, 
PRONEX and FAPERJ (Brazilian agencies) is also acknowledged.

\bigskip

{C. Tsallis$^{1,2}$, B.J.C. Cabral$^1$, A. Rapisarda$^3$ and V. Latora$^3$

\bigskip
\noindent
$^1$Centro de Fisica da Materia Condensada,
 Universidade de Lisboa,Av. Prof. Gama Pinto 2, P-1649-003 Lisboa, Portugal

\noindent
$^2$Centro Brasileiro de Pesquisas Fisicas, 
Rua Xavier Sigaud 150,
  22290-180 Rio de Janeiro-RJ, Brazil

\noindent
$^3$Dipartimento di Fisica e Astronomia,   Universit\'a di Catania,
and INFN, Sezione di Catania, Corso Italia 57, I-95129 Catania, Italy
}

\bigskip
\noindent
{$^*$ Email: tsallis@cbpf.br, ben@adonis.cii.fc.ul.pt, 
andrea.rapisarda@ct.infn.it,
 vito.latora@ct.infn.it}


\begin{figure}
\begin{center}
\epsfig{figure=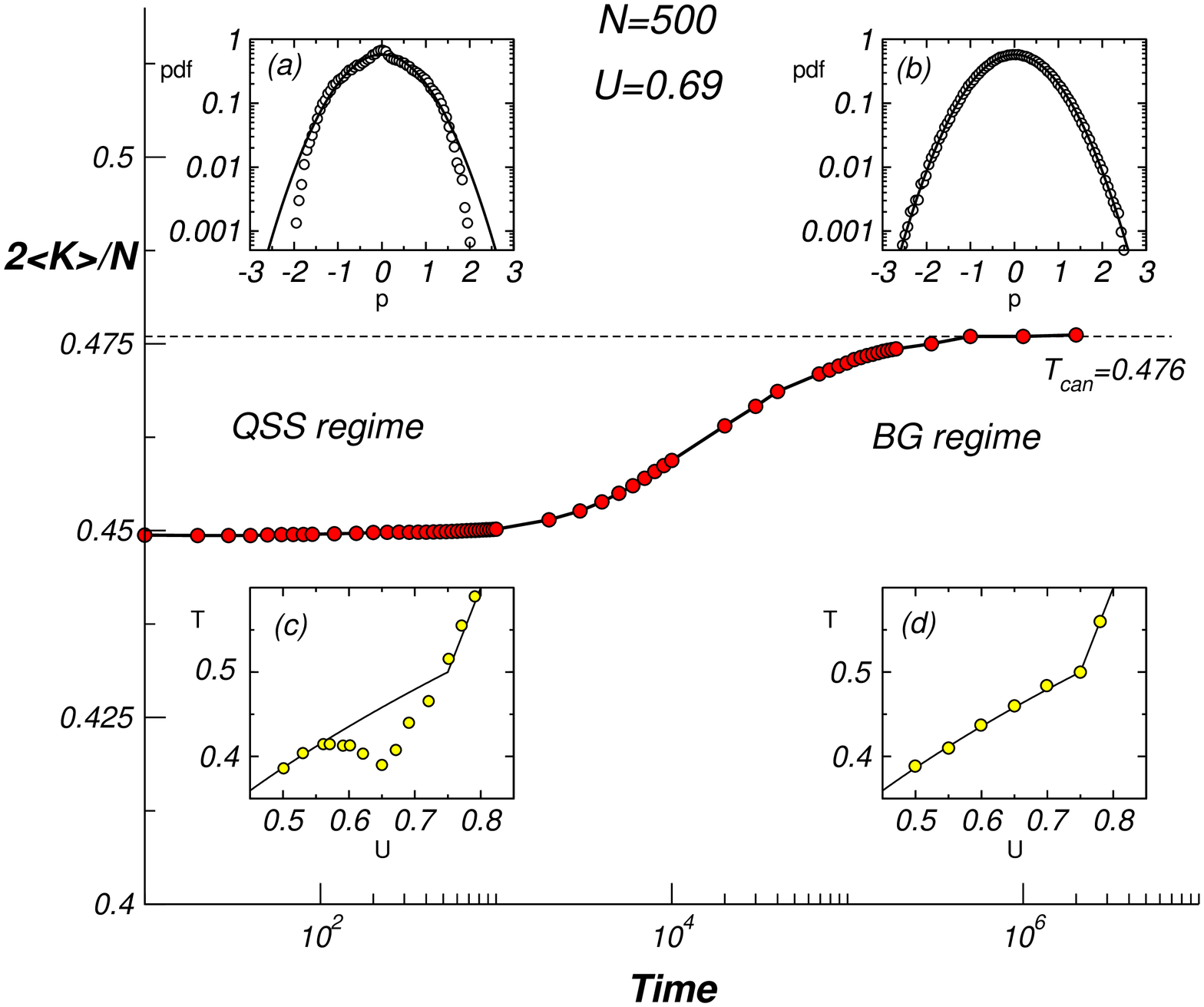,width=14truecm,angle=-90}
\end{center}
\caption{ 
Time evolution of twice the average kinetic energy per particle
 (``temperature") for a HMF system with N=500 (red circles) at 
energy per particle $U=0.69$.
 The corresponding velocity distribution is {\it not 
Maxwellian} for the  first, metastable  plateau 
(panel (a)), whereas
 standard statistical mechanics is  valid for the second,
 stable plateau (panel (b)), open circles. 
The Gaussian canonical prediction
 is also shown as a full curve.  Correspondingly, we find a negative 
specific heat, $C_V=dU/dT$, in the 
metastable state (Quasi-Stationary State (QSS)
regime) , as shown in panel (c)
 (open circles) in comparison with the canonical caloric curve, while
the specific heat for this hamiltonian  is always positive when 
BG equilibrium is reached (panel (d)).
Such anomalies  are  strongly related to  nonextensivity 
and have been found in various systems \protect\cite{borges,sota,hmf2,nonmaxwell}.
}
\end{figure}
\end{document}